\documentclass[aps,prl,preprint,showpacs,groupedaddress]{revtex4}

\usepackage {amssymb}
\usepackage {amsmath}
\usepackage {graphicx}
\usepackage {longtable}
\usepackage {color}

\begin{document}
\title{Mixed (1D-2D) quantum electron transport in percolating
gold film}

\author{E. Yu. Beliayev}
\affiliation{B. Verkin Institute for Low Temperature Physics and
Engineering, National Academy of Sciences, Kharkov 61103, Ukraine}

\author{B. I. Belevtsev}
\email[]{belevtsev@ilt.kharkov.ua}
\affiliation{B. Verkin Institute for Low Temperature Physics and
Engineering, National Academy of Sciences, Kharkov 61103, Ukraine}

\author{Yu. A. Kolesnichenko}
\affiliation{B. Verkin Institute for Low Temperature Physics and
Engineering, National Academy of Sciences, Kharkov 61103, Ukraine}

\begin{abstract}
The gold film (mean thickness $\approx$ 3.5 nm) was condensed in
high vacuum at temperature 70 K on single-crystal sapphire
substrate. The transport properties of the film at low temperature
reveal simultaneously indications of 1D and 2D quantum
interference effects of weak localization and electron-electron
interaction. It is shown that this behavior is determined by
inhomogeneous electron transport at the threshold of
thickness-controlled metal-insulator transition.
\end{abstract}
\pacs{73.20.Fz, 73.50.Jt, 71.30.+h}
 \maketitle

\section{Introduction}
\label{int} Metal-insulator transition (MIT) in disordered systems
is still a basic challenge in the solid-state physics. One of the
principal causes of MIT is a sufficient increase in lattice
disorder (so called Anderson transition). It was theoretically
stated that a two-dimensional (2D) system of non-interacting
electrons should be insolating at any degree of disorder when
going to zero temperature \cite{abraham}. Electron-electron
interaction plays, however, crucial role in 2D transport
especially at low temperature \cite{altsh1,gant,altsh2}, so that
MIT in 2D systems is asserted theoretically as possible which
finds also some experimental support \cite{altsh1,gant}. Two types
of crystal-lattice disorder affecting electron propagation in
conducting solids are usually considered. The first is associated
with perturbations of the scattering potentials on the atomic
scale: impurities, vacancies, and so on, and is often called
microscopic. At the same time, MIT can also take place in
inhomogeneous systems such as disordered mixtures of a metal and
an insulator, granular metals, and the like
\cite{belufn,beloborod}, where the scale of disorder is well above
interatomic distances, and where propagating electrons should
overcome insulating regions (or boundaries) between the metallic
clusters (grains). This second type of disorder is called
macroscopic. MIT in systems with macroscopic disorder is
inevitably associated with (classical or quantum) percolation
effects in electron transport.
\par
A good few known experimental studies of MIT have been performed
with ultrathin metal films by means of step-by-step increasing of
the film thickness (see reviews \cite{gant,belufn} and references
therein). The nominal thickness increments at these studies are
usually in the range 0.05-0.1~\AA, that is much less than
interatomic distances. This brings the questions: are the films
remain homogeneous in thickness at these small increments, and are
the films homogeneous at all at MIT, which occurs at some critical
thickness, $d_c$? It is known from available literature about this
type of MIT studies, that $d_c$ is in the range 1--5 nm depending
on film structure (crystalline or amorphous) and material of
substrate. The minimum values of $d_c$ (1--1.5 nm) were found in
quench-condensed amorphous Bi and Ga films with thin underlayer of
amorphous Ge \cite{gold}. For the most cases the sheet resistance
of films, $R_{\square}$, with thickness $d\approx d_c$ is order of
10 k$\Omega$, which is comparable with the resistance quantum
$R_Q=R_{\square} =a\hslash/e^2$, where $\hslash/e^2\approx 4.1
$~k$\Omega$ and $a$ is order of unity. $R_Q$ is often considered
as some characteristic resistance for 2D MIT. It is evident that
condensed films with $R_{\square}\geq R_Q$ are discontinuous, so
that percolating effects should be important.
\par
In this study, we present transport properties of quench-condensed
gold film with thickness about 3.56 nm and $R_{\square}$ about 5
k$\Omega$ above 10 K. This film has weak nonmetallic temperature
dependence of resistance with logarithmic behavior above 10 K and
somewhat stronger dependence at low temperature. The film is
characterized by quantum interference effects in electron
transport. Above 3 K only 2D effects in transport have been found;
whereas, below 3 K both one-dimensional (1D) and 2D effects in
transport properties can be distinguished. This reflects
inhomogeneous structure (and corresponding electron transport) of
the film near the thickness-controlled MIT in ultrathin metallic
films.

\section{Experimental Technique}
The gold film studied has been prepared by thermal evaporation
from a Mo boat onto a substrate of polished single-crystal
sapphire plate. The size of the film regions being measured was
$2\times 0.1$ mm$^2$. The substrate was preliminarily etched in
hot aqua regia, then washed in distilled water and etched in
boiling bichromate, then again washed in distilled water, and
finally dried in dessicator.
\par
Deposition of the film on the prepared in this way substrate and
subsequent measurement {\it in situ} its transport properties were
carried out in the high-vacuum cryostat containing $^3$He and a
superconducting solenoid. Oil-free vacuum pumps were used to reach
pressure about $1\times 10^{-7}$ Pa. A system of sliding screens
(masks) was used for deposition of  gold contacts (at room
temperature) and subsequent deposition of the gold film (at
$T\approx 70$ K with the rate 0.05 nm/s). Initial purity of gold
deposited was 99.99\%. Pressure of residual gas during deposition
was about $5\times 10^{-6}$ Pa. Resistance of the film during its
growth has been followed (under applied voltage $U_{appl}=1$ V),
and deposition was interrupted when $R_{\Box}$ had reached about
4.4 k$\Omega$. According to our previous studies \cite{belev1},
structure  of gold films with $R_{\Box}$ about 5 k$\Omega$
deposited at similar conditions is close to percolation threshold.
The nominal film thickness, determined by a quartz sensor, is 3.56
nm. The sample prepared was held at the preparation temperature
(70 K) during 12 h up to stopping of small resistance variation
due to structural relaxation inherent for films deposited on cold
substrates. The magnetoresistive measurements were carried out
{\it in situ} in lower temperature range 0.4--30 K in magnetic
fields from -0.05 to +5 T. The resistance was determined by
measuring current in conditions of preset applied voltage
$U_{appl}$. The most of measurements have been done for
$U_{appl}=$ 0.2, 1 and 5 V, although other voltages were also
used, especially for determination of the current-voltage
characteristic at different temperatures (see below).

\section{Results and discussion}
\subsection{General characterization of transport properties of the film}
It is found that conducting state of the film studied corresponds
to that on the threshold of the thickness controlled MIT.
Temperature dependence of resistance has nonmetallic ($dR/dT <0$)
character, which above 10 K and at high enough applied voltage
follows the logarithmic law ($\Delta R \propto \ln T$) (Fig.~1).
In the low temperature range (below 5 K), the explicit dependence
of resistance on applied voltage takes place (Figs. 1 and 2). It
is seen (Fig. 2) that at $T=5$~K a decrease in voltage leads only
to a slight increase in resistance so that at this temperature
(and above it) Ohm's law is obeyed to a good approximation. At
lower temperature, however, the exponential increase in resistance
with decreasing voltage is seen in low-voltage range ($U_{appl}<
50$~mV) (inset in Fig. 2), although for $U_{appl}\geq 2$~V the
non-Ohmic behavior is pronounced insignificantly. It should be
noted that the same voltage dependence [$R\propto
\exp(1/U_{appl})$] in low-temperature range (but with far larger
amplitude in $R$ variations) has been observed in Ref.
\cite{belev1} in percolating gold film obtained under the same
conditions as in this study. But being with less nominal thickness
(3.25 nm) the film in Ref. \cite{belev1} was accordingly more
resistive and demonstrated a transition from strong to weak
electron localization under increasing $U_{appl}$ in full measure;
whereas, in the film studied only some indications of such type of
behavior can be found.
\par
Thus the transport properties testify to inhomogeneous structure
of the film. This is sure connected with spatial inhomogeneity in
thickness. The small effective thickness of the film (3.56 nm),
high values of $R_{\Box}$ (about 6 k$\Omega$ and higher) and
rather strongly pronounced nonohmic behavior of its conductivity
in low temperature range indicate that the film is partially
discontinuous. This means that it consists of metallic islands (or
percolating clusters) separated by tunnel barriers. Two types of
barriers are possible: (a) vacuum gaps between adjacent islands on
an insulating substrate, and (b) narrow and thin constrictions
(bridges) between islands in the case of weak contacts between
neighboring islands. In any case, such a system can be regarded as
similar to a two-dimensional granular metal. These suggestions are
fully in line with the results of known studies of
quench-condensed (substrate temperature between 4 K and 77 K) Au
films \cite{ekinci}. It was found in particular \cite{ekinci} that
the critical nominal thickness, $d_{cr}$, marking the onset of
conductivity is about 2 nm for quench-condensed Au films that
agrees reasonably well with results of this study. It follows from
results of Ref. \cite{ekinci} that quench-condensed Au films
deposited on a weak-binding substrate (like that in this study)
form a 2D disordered array of weak-connected islands at
thicknesses moderately greater than $d_{cr}$. The grain diameter,
$d_G$, in this type of Au film is in the range
10~$\textrm{nm}<d_G<20$~nm \cite{ekinci}.
\par
A quasi-granular 2D structure of the film is also supported by the
following consideration. It is known that product $\rho l$ [where
$\rho$~($=R_{\Box} d$) and $l$ are the resistivity and the
mean-free path, respectively] is characteristic constant for
typical metals. For gold, $\rho l\approx 8.43\times
10^{-16}$~$\Omega  \textrm{m}^2$ can be found in the framework of
the quasi-free electron approximation. Taking  this relation,
$l\approx 0.05$~nm can be obtained for the film studied. So low
nominal value (less than interatomic distance) indicates clearly
an inhomogeneous structure of the film studied on the threshold of
the thickness controlled MIT. The high resistivity is determined
by low intergrain tunnel conductivity; whereas, intragrain
conductivity can be far larger. We will take into account the
particular morphology of the quench-condensed ultrathin gold films
at discussion of the results obtained.

\subsection{Quantum interference effects in conductivity of the
film studied}
\par
The logarithmic law ($\Delta R \propto \ln T$) found in thin film
studied (Fig. 1) implies that it is connected with quantum weak
localization (WL) and electron-electron interaction (EEI) effects
in conductivity of 2D systems \cite{altsh2}. The two-dimensional
conditions of manifestation of these effects are: $d <
L_{\varphi}, L_{T}$, where $d$ is the film thickness,
$L_{\varphi}$ = $(D\tau_{\varphi})^{1/2}$ is the diffusion length
of phase relaxation, $L_{T} = (\hbar D/kT)^{1/2}$ is the thermal
coherence length in a normal metal, $D$ is the electron diffusion
coefficient, and $\tau_{\varphi}$ is the time of phase relaxation.
The lengths $L_{\varphi}$ and $L_{T}$ are attributed to the WL and
EEI effects, respectively. Macroscopic disorder (percolating
and/or granular structures) can induce a dramatic influence on the
WL and EEI effects \cite{palev} up to their total depression at
strong enough disorder. At the same time, a rather weak
macroscopic disorder has no significant influence on these quantum
effects. In particular, a system behaves as homogeneous in respect
to WL and EEI effects if the relevant lengths [$L_{\varphi}(T)$
and $L_T$] are larger than characteristic inhomogeneity scale (for
example, percolation correlation length, $\xi_p$, or grain
diameter, $d_G$, in 2D granular metal).
\par
For homogeneous 2D system the contribution of WL and EEI to the
temperature dependence of conductivity in zero magnetic field is
given by \cite{altsh2}
\begin{equation}
\Delta \sigma (T) = \frac{e^{2}}{2\pi^{2}\hbar}\left\{-\left
[\frac{3}{2}\ln \frac{\tau_{\varphi}^{\ast}}{\tau} -
\frac{1}{2}\ln \frac{\tau_{\varphi}}{\tau}\right ] +
\lambda_{T}^{D}\ln \frac{kT\tau}{\hbar} \right\} \label{one}
\end{equation}
where $\tau$ is the elastic electron relaxation time,
$\tau_{\varphi}^{-1} = \tau_{in}^{-1} + 2\tau_{s}^{-1}$;
$(\tau_{\varphi}^{\ast})^{-1}  = \tau_{in}^{-1} +
(4/3)\tau_{so}^{-1} + (2/3)\tau_{s}^{-1}$, $\tau_{in}$ is the
phase relaxation time due to inelastic scattering, $\tau_{so}$ and
$\tau_{s}$ are spin relaxation times of electrons due to
spin-orbit and spin-spin scattering, respectively.
$\lambda_{T}^{D}$ is the interaction constant in the diffusive
channel, which is  close to unity for typical metals. The time
$\tau_{in}$ is temperature dependent, namely, $\tau_{in}^{-1}
\propto T^{p}$, where $p$ is the exponent which depends on the
mechanism of inelastic scattering. The first term in Eq.(1)
corresponds to WL effects, while the second one to EEI.
\par
The Eq.\ (\ref{one}) presents well known logarithmic correction to
conductivity  which can be rewritten as \cite{altsh2}
\begin{equation}
\frac{\Delta R}{R} = - a_{T}\,
\frac{e^{2}R_{\Box}^{eff}}{2\pi^{2}\hbar}\: \ln\left
(\frac{kT\tau}{\hbar}\right ), \label{two}
\end{equation}
\noindent with $a_T$ order of unity, exact value of which is
determined by dominating mechanisms of phase relaxation in a
specific system.

\par
It is known that gold is characterized by strong spin-orbit
interaction [$\tau_{so} \ll \tau_{in}(T)$], which determines
specific value of $a_T$, and causes appearance of positive
magnetoresistance due to the WL effect \cite{altsh2}, as it is
usually observed in gold films (Refs. \cite{belev1, belev2} and
Refs. therein), including the film studied (see Fig. 1 and more
detailed information below). From Eq.\ (\ref{one}), on conditions
of strong spin-orbit scattering and neglecting the spin-spin
interaction [$\tau_{so} \ll \tau_{in}(T)$ and $\tau_{s} \gg
\tau_{so}, \tau_{in}$], it can be obtained that $a_{T} = 1 - p/2$,
where unity comes from the EEI effect and second part is
determined by the WL effect with $p$ being the exponent in
temperature dependence $\tau_{in}^{-1}\propto T^p$ for a dominant
mechanism of phase relaxation.
\par
Generally, the phase relaxation is determined by the two main
contributions: electron-electron and electron-phonon interactions,
so that $\tau_{in}^{-1}= \tau_{ee}^{-1}+\tau_{ep}^{-1}$, where
$\tau_{ee}$ and $\tau_{ep}$ are the corresponding times. It was
found for the electron-phonon interaction processes in disordered
($R_{\Box}$ up to $\approx 500$~$\Omega$) gold films that
$\tau_{ep}^{-1}\propto T^{p}$ with exponent $p$ between 2 and 2.8
\cite{belev2}. According to Ref. \cite{belev2}, the
electron-phonon scattering dominates the rate of phase relaxation
in gold films only above 10 K. Below this temperature
$\tau_{in}^{-1}\propto T$ was found and attributed to influence of
electron-electron scattering. It is known that the phase
relaxation rate due to electron-electron scattering in 2D systems
is determined by collisions with small energy transfer ($\Delta
E\ll kT$) \cite{altsh2}
\begin{equation}
\tau_{ee}^{-1}(T) = \frac{\pi kT}{\hbar}\:
\frac{e^{2}R_{\Box}}{2\pi^{2}\hbar}\: \ln \left (\frac{\pi
\hbar}{e^{2}R_{\Box}}\right ) \label{three}
\end{equation}
\noindent Increasing disorder leads to domination of EEI in the
phase relaxation (that is $\tau_{ee}^{-1} \gg \tau_{ep}^{-1}$).
Since the film studied is far more disordered than those in Ref.
\cite{belev2} the temperature range of EEI domination in it should
be much wider. In this case, $p = 1$ should be taken and,
consequently, $a_{T} = 1/2$ is to expect.
\par
Now let us consider experimental values of $a_T$. It is found in
the film studied that $a_T\approx 1.15$ at $U_{appl}= 0.2$~V at
$T>8$~K and it is close to unity for higher voltage in the range
above 5 K (Figs. 1 and 3). When using Eq.~(\ref{two}) for deriving
$a_T$ values we have taken the measured macroscopic resistance
$R_{\Box}^{exp}$ as $R_{\Box}^{eff}$. This may be done only for
homogeneous systems. If a system can be supposed to be
inhomogeneous, $R_{\Box}^{eff}$ should be a fitting parameter
\cite{gersh}. We have checked out this matter. For instance, from
analysis of logarithmic $R(T)$ dependence at $U_{appl}= 5$~V we
found that $R_{\Box}^{eff}=5.28$~k$\Omega$ given $a_T=1$. This
sheet resistance value falls within range of the $R_{\Box}$
temperature variations  at $U_{appl}= 5$~V (Fig. 1), so that at
this voltage the system behaves as homogeneous one above 5 K.
\par
The derived values of $a_T\approx 1$ do not correspond to
$a_T\approx 1/2$ expected for systems with strong spin-orbit
interaction. So it looks like that only EEI makes contribution to
logarithmic $R(T)$ dependence. A rather plain suggestion could be
that in the film studied the WL correction is significantly
suppressed for some reasons so that EEI contribution has dominant
effect on $R(T)$ which results in $a_{T}\approx 1$. The total
suppression of WL is possible in some cases, for example, by
application of strong enough magnetic field ($H> H_{\varphi}=\hbar
c/4eL_{\varphi}^2$) \cite{altsh2}. As it is seen from Fig.~1, the
application of field $H=4.2$~T, which is far larger than
$H_{\varphi}$ (as will be clear below) causes only weak effect on
a slope of the linear part of dependence $\Delta R$ {\it vs} $\ln
T$ above $T\approx 8$~K, so that it seems that WL is significantly
suppressed in the film even in zero field. This situation is
possible in inhomogeneous (percolating or granular) 2D films
\cite{beloborod}. In a system consisting of rather large grains
separated by weak tunnel barriers appearance of the WL correction
(same as in an homogeneous system) is determined by closed
electron trajectories with self-intersections that increases the
probability of back-scattering \cite{altsh2}. The size of these
trajectories is about $L_{\varphi}$. The EEI correction is
determined by length $L_T$. Generally, in the case of
quasi-particle description of electrons, the relation $L_T\ll
L_{\varphi}$ holds \cite{altsh2}. In any case, for the WL and EEI
corrections to be clearly seen in $R(T)$ dependences, the both
lengths $L_{\varphi}$ and  $L_T$ should exceed the characteristic
inhomogeneity scale of the system (for example, grain size $d_G$).
Since $L_{\varphi}\gg L_T$, electron trajectories which determine
WL effect should include more grains (and more intersections with
grain boundary) than those connected with EEI effect, so that weak
intergrain connections can induce more depressing effect on WL
correction than that on EEI one.
\par
It is seen in Fig. 1 that $R(T)$ follows the logarithmic law only
for high enough temperature.  $R(T)$ begins to deviate from $\ln
T$ behavior below 5 K at $U_{appl}=5$~V, and seems to go to
saturation at low enough temperature. In this case an overheating
effect cannot be excluded. The overheating should be, however,
diminished with decreasing voltage. Really, the saturation has
disappeared at $U_{appl}=1$~V, and for the lower voltage
$U_{appl}=0.2$~V even an increased rate of the $R(T)$ growth with
decreasing temperature is observed in low temperature range
(Fig.~1). The latter case is presented more clearly in Fig. 3.
Below 9 K a deviation from $\ln T$-law appears, which is presented
in the inset as $\Delta R = f(T^{-1/2})$. This will be discussed
in detail below.
\par
The magnetoresistance (MR) data obtained reflect inhomogeneous
structure of the film studied as well. General view of MR curves
and their variations with temperature can be seen in Fig. 4.
Except for low-field ($H\leq 0.02$~T) and low-temperature ($T\leq
3.1$~K) ranges (see below), the magnetoresistive curves are found
to correspond to the known expression describing MR due to WL
effect for 2D systems \cite{altsh2}:
\begin{equation}
\Delta \sigma (H) = \frac{e^{2}}{2\pi^{2}\hbar}\left\{
\frac{3}{2}f_{2}\left (\frac{4eDH\tau_{\varphi}^{\ast}}{\hbar
c}\right ) - \frac{1}{2}f_{2}\left
(\frac{4eDH\tau_{\varphi}}{\hbar c} \right ) \right\} \label{four}
\end{equation}

\noindent where $f_{2}(x) = \ln (x) + \Psi (1/2 + 1/x)$, $\Psi$ is
digamma function. $f_{2} (x)\approx x^2/24$ for $x\ll 1$, and
$f_{2}(x)\propto \ln x$ for $x\gg 1$.  For the films studied, the
contribution of EEI to MR is negligible, so we do not cite
corresponding expressions.
\par
Positive MR found (Fig. 4) is expected for gold films with strong
spin-orbit scattering. For this case [$\tau_{so} \ll
\tau_{in}(T)$], Eq. (\ref{four}) can be rewritten as \cite{altsh2}
\begin{equation}
\frac{\Delta R(H)}{R} = R_{\Box}^{MR}\frac{e^{2}}{4\pi^{2}\hbar}
f_{2}\left (\frac{4eHL_{\varphi}^{2}}{\hbar c} \right ).
 \label{five}
\end{equation}
We have found that Eq. (\ref{five}) describes adequately the
experimental $R(H)$ dependences for $H\leq$~1~T with small
deviations for higher field. The Eq. (\ref{four}) describes well
experimental $R(H)$ in the whole field range, and the both
equations have been used to derive the $L_{\varphi}(T)$
dependence.
\par
For homogeneous systems, measured macroscopic resistance
$R_{\Box}^{exp}$ can be substituted for $R_{\Box}^{MR}$ when
processing experimental data by Eqs. \ref{four} or \ref{five}.
$R_{\Box}^{MR}$ is resistance on the scale $L_{\varphi}$, which in
an inhomogeneous sample can be less than macroscopic resistance
measured on much larger scale. In this case, $R_{\Box}^{MR}$
should be a fitting parameter \cite{gersh,but}. We have really
found that computed values of $R_{\Box}^{MR}$ are somewhat less
than $R_{\Box}^{exp}$ (on about 20\%). This difference is not,
however, crucial and we have not found a significant effect of it
on calculated values of $L_{\varphi}$.
\par
The temperature dependence of $L_{\varphi}(T)$ in the range
1.4--30 K is shown in Fig. 5. It is seen that generally
$L_{\varphi}(T)$ is not essentially dependent on the applied
voltage. In particular, in the range 3-10 K, the data for
different $U_{appl}$ practically coincide. Above $T\approx 10$~K
the data become more noisy due to smaller magnitude of MR in this
range. In low-temperature range (below 3 K), influence of
$U_{appl}$ on $L_{\varphi}$ is evident. In this range the
$L_{\varphi}(T)$ values tend to saturation with decreasing
temperature below 3 K. The following temperature dependences
$L_{\varphi}(T)$ were found: $L_{\varphi}(T)\propto T^{-0.35}$
(3--15 K) and $L_{\varphi}(T)\propto T^{-1.05}$ (15--30 K). Since
$L_{\varphi}\propto T^{-p/2}$, the first case corresponds to
$p\approx 0.7$ and the second to $p\approx 2.1$. We will stop at
these features of $L_{\varphi}(T)$ in the next section.

\subsection{Enhancement of percolating character of electron transport
at low temperature and 1D  effects in the film conductivity}
\par
In low temperature range we have found a weak but quite distinct
anomaly of MR behavior in low-field range, which is shown on a
large scale in Fig. 6. This anomaly disappears with increasing
field  and is depressed with increase in temperature up to
$T\simeq 3.1$~K (Fig. 7). The field at which the MR anomaly comes
to saturation (and to transition to 2D MR behavior) increases
with decreasing temperature being in the range 0.01-0.03 T (Fig.
7). At low field ($H\leq 0.004$~T) MR in the range of this anomaly
follows the quadratic law $\Delta R(H)/R(0)\propto H^2$ (Fig. 8).
Temperature behavior of maximum MR of the anomaly (at field
position of saturation) is shown in the inset of Fig. 8.
\par
It should be noted that this type of anomaly have been seen by us
more than once at low enough temperature in other fairly
disordered gold films (close to the thickness controlled
percolation threshold). Results obtained permit us to assert that
this low-temperature MR anomaly together with the above-mentioned
deviations of $R(T)$ from logarithmic behavior at low temperature
range (Fig. 3) are indications of 1D effects in film conductivity
caused by inhomogeneous (granular) film structure near the
percolation threshold. Indications of mixed 1D-2D conductivity in
percolating gold films were first suggested to be seen in Ref.
\cite{palev}. Further 1D interference effects had been studied in
specially made narrow enough films (see
\cite{gersh,wind,willi,gersh2} and references therein). The theory
of quantum interference (WL and EEI) effects in 1D conductors has
been developed in Refs. \cite{altsh2,altsh3,altsh4}.
\par
Before doing further analysis and discussion of the 1D effects in
the film studied, the principal causes of appearance of these
effects in ultrathin films should be considered. As was mentioned
above, a film near the thickness-controlled percolation threshold
consists of weakly connected islands (grains). The intergrain
connections are determined by narrow constrictions, which can make
tunnel junctions (like point contacts). In real quasi-island films
near the percolation threshold the intergrain constrictions
(contacts) are not the same throughout the system, thus, the
conductivity is percolating. It is determined by the presence of
"optimal" chains of grains with maximum probability of tunnelling
for adjacent pairs of grains forming the chain. At low enough
temperature the tunnelling can become activated. In conditions of
activated conductivity, the number of conducting chains decreases
with decreasing temperature, so that at low enough temperature a
percolation network can consist of a few conducting channels or
even come to a single conducting path \cite{sheng}. It appears to
be a general trend for disordered systems, and in particular for
2D systems. For example, it was theoretically shown \cite{mark},
that increasing in disorder in 2D system leads to forming of a
narrow channel along which electrons propagate through a
disordered sample.
\par
It follows from the aforesaid that thin enough percolating film
can manifest a mixed (1D and 2D) behavior of the interference
effects in the conductivity. In this case, similar as in Ref.
\cite{palev}, it can be suggested that quantum correction to
conductivity has two contributions,
\begin{equation}
\frac{\Delta R(T,H)}{R}= \left (\frac{\Delta R(T,H)}{R} \right
)_{1D} + \left (\frac{\Delta R(T,H)}{R} \right )_{2D},
 \label{six}
\end{equation}
where lower indices $1D$ and  $2D$ mark the corresponding
contributions. Relative contribution of each term is temperature
and magnetic-field dependent, so that, for example, the 1D term
can even disappear at high enough  $T$ or $H$.
\par
Expressions for the 2D corrections have been presented above. For
further discussion some valid 1D expressions should be considered.
Conditions for appearance of 1D interference effects in electron
transport are that both characteristic lengths, $L_T=(\hbar
D/kT)^{1/2}$ and $L_{\varphi}=(D\tau_{\varphi})^{1/2}$, must be
greater than width, $W$, and thickness, $d$, of the 1D wire. It is
known \cite{altsh2,wind,gersh2} that 1D effects become apparent
only at low temperatures where the main contribution to the phase
breaking gives the electron-electron scattering with small energy
transfer \cite{altsh2,altsh3,altsh4}. This is, so called, Nyquist
phase-breaking mechanism which is especially important for
low-dimensional systems \cite{altsh2,altsh4,gersh2}. For 2D
systems the Nyquist time is presented by Eq.~(\ref{three}). For 1D
systems this time is denoted by $\tau_N$, and corresponding phase
relaxation rate is given by \cite{altsh2,altsh4}
\begin{equation}
\tau_{N}^{-1}= \left [\left (\frac{e^{2}R_{\Box}}{\hbar}\right )
\left (\frac{kT}{\hbar}\right ) \left (\frac{\sqrt{2D}}{W}\right )
\right ]^{2/3} \propto T^{2/3}.
 \label{seven}
\end{equation}
The characteristic phase-relaxation length for this process is
$L_N= (D\tau_N)^{1/2}\propto T^{-1/3}$.
\par
In line with our suggestion about mixed (1D and 2D) behavior of
the film conductivity at low temperature we have compared the
deviations of $R(T)$ from logarithmic 2D behavior at low
temperature range (Fig. 3) with known 1D expressions. EEI
correction to the resistance a single film strip (at $W< L_T$)
is \cite{altsh2}
\begin{equation}
\frac{\Delta R_{int}(T)}{R}= \frac{e^2}{2\hbar} \
\frac{R_{\Box}^{str}}{W} \ L_T \propto T^{-1/2},
 \label{eight}
\end{equation}
where $R_{\Box}^{str}$ is sheet resistance of the strip. This
quantum correction increases with decreasing temperature as
$T^{-1/2}$, that agrees rather well with the $\Delta R (T)$
behavior shown in inset of Fig. 3. It should be noted that
dependence $\Delta R \propto T^{-1/2}$ in the film studied takes
place also at high magnetic field which depressed completely WL
effect, so that this behavior should be attributed solely to EEI
effect. Similar behavior corresponding to Eq. (\ref{eight}) has
been previously seen in 1D Au films \cite{gersh,willi}.
\par
Of course, WL effect can also influence quantum 1D transport
\cite{altsh2,altsh4}. However, since gold is characterized by
strong spin-orbit scattering (antilocalization), the corresponding
correction, $\Delta R_{loc}(T)$,  is negative and $R$ should
decrease with decreasing temperature. It was found in 1D gold
films that at low enough temperature EEI effect gives dominating
contribution to the quantum correction \cite{gersh,gersh2}.
Apparently this occurs in the gold film studied as well.
\par
Consider now in more detail the low-field anomaly of MR in
low-temperature range (Figs. 6 and 7), which we have attributed to
1D effects as well. MR of 1D systems in low magnetic field is
determined exclusively by WL \cite{altsh2}. At low enough
temperature the phase relaxation is determined by Nyquist
mechanism so that $\tau_N\ll\tau_{\varphi 0}(H=0)$, where
$\tau_{\varphi 0}$ is phase relaxation time  due to all other
phase-breaking mechanisms. In the case of strong spin-orbit
scattering, WL correction to resistance of 1D film for magnetic
field perpendicular to the film plane is
\cite{altsh2,gersh2,altsh3}
\begin{equation}
\frac{\Delta R_{loc}(T,H)}{R_0}= -0.31\ \frac{e^{2}R_{\Box}^{str}}{\hbar
W}\ [1/L_{N}^{2}+1/D\tau_H]^{-1/2},
 \label{nine}
\end{equation}
where $\tau_H=12L_{H}^{4}/DW^{2}$ and $L_H=\sqrt{\hbar c/2eH}$. It
is seen from Eq. (\ref{nine}) that at low fields ($L_{N}^{2}\ll
D\tau_{H}$) increase in resistance in field $H$ is
\begin{equation}
\frac{\Delta R_{loc}(T,H)- \Delta R_{loc}(T,0)}{R_0}\approx
\frac{0.31}{2}\ \frac{e^{2}R_{\Box}^{str}}{\hbar W}\
\frac{L_{N}^{3}}{D\tau_H}= \frac{0.31}{2}\
\frac{e^{2}R_{\Box}^{str}}{\hbar}\
\frac{L_N^{3}W}{12L_H^{4}}\propto H^2/T.
 \label{ten}
\end{equation}
Quadratic dependence of MR on magnetic field for low-field range
is clearly seen (Fig. 8).  The temperature dependence of MR is
difficult to determine exactly enough due to narrow temperature
range of the anomaly. We have found, however, that this dependence
is strong (inset in Fig. 8) and in the range 1.4--3 K follows (but
rather approximately) the $1/T$ law, as can be expected. It should
also be mentioned that the amplitude of MR attributed to 1D WL
found in this study ($\Delta R(H)/R$ of the order of $10^{-4}$)
agrees well with those found for other 1D metal films, including
Au films \cite{gersh2,lin,her,pier}.
\par
Quantitative comparison with Eq. (\ref{ten}) and obtaining values
of $L_N$ from it appears to be quite difficult (or even, at first
glance, practically impossible) since we actually do not know
exact values of two parameters: sheet resistance,
$R_{\Box}^{str}$, of 1D strip and its width W. Macroscopic film
resistance $R_{\Box}^{exp}\approx 5.2$~k$\Omega$ can be
substituted for $R_{\Box}^{str}$, but this can cause some error in
calculated values of $L_N$ since it is expected that
$R_{\Box}^{str} \ll R_{\Box}^{exp}$. A rough estimate of $W$ can
be done taking into account that in 1D films the transition from
the 1D to 2D WL behavior of MR takes place when $L_H$ with
increasing field becomes smaller than width $W$ \cite{gersh2}. It
follows from Fig.~7 that in the film studied the transition field,
$H_{tr}$, at which the MR anomaly saturates, depends on
temperature. The magnitude of $H_{tr}$ grows from $\approx 0.013$
T to $\approx 0.03$~T with decrease in temperature from $T=3$~K to
$T=0.38$~K, which implies decreasing in $W$ from $\approx 160$~nm
to $\approx 100$~nm for this temperature variation (if to take
magnetic lengths $L_H$ for these values of $H_{tr}$ as values of
$W$). Decreasing in $W$ with decreasing temperature is not
surprising and, furthermore, quite expected in the light of the
above-discussed reasons and conditions of forming of 1D
percolation structure in quench-condensed ultra-thin films.
\par
Taking nominally the obtained in the above-mentioned way values of
$W$ and  $R_{\Box}^{exp}\approx 5.2$~k$\Omega$, we have calculated
$L_N$ values at different temperatures with help of
Eq.~(\ref{ten}) for low-field MR. The result is shown in Fig. 9.
It can be seen that in the range 1.5--3 K the calculated values of
of $L_N$ are proportional to $T^{-0.386}$ which is close enough to
theoretical prediction $L_N=(D\tau_N)^{1/2}\propto T^{-1/3}$.
Derived values of $L_N$ appear to be rather larger than those of
length $L_{\varphi}=(D\tau_{\varphi})^{1/2}$, determined from 2D
MR curves in the same range 1.4--3 K (compare Figs. 5 and 9). It
is important to mention as well that obtained values of $L_N$
exceed the characteristic inhomogeneity scale of the granular film
studied, the grain size, which, according to Ref. \cite{ekinci},
is between 10 and 20 nm in quench-condensed ultrathin gold films.
This is the necessary condition for observation of WL effects in
granular or island films. Unfortunately, it is not possible to
find the magnitudes of corresponding relaxation times, $\tau_N$
and $\tau_{\varphi}$ since we cannot determine true values of
respective electron diffusion coefficients $D$ for the 1D and 2D
cases in the inhomogeneous percolating film studied.
\par
It should be noted that calculated values of $L_N$ are quite
smaller than estimated width W of inferred 1D strip(s) in the film
studied. As it is indicate above, this can be connected with
uncertainty in value of $R_{\Box}^{str}$ in processing data with
Eq. (\ref{ten}): $R_{\Box}^{str}$ can be even ten times smaller
than the value which is used (5.2 k$\Omega$). The uncertainty in
$R_{\Box}^{str}$ should not, however, affect appreciably the
temperature dependence $L_{N}(T)$, which we have obtained
(Fig.~9).
\par
For both, 1D and 2D WL interference effects in the film studied,
the phase relaxation lengths obtained seem to saturate going to
low temperature (Figs. 5 and 9). The saturation is more pronounced
for higher $U_{appl}$ (Fig. 5) that can be determined by
overheating effect. But the saturation persists even at lower
$U_{appl}$, and this can be determined by 2D inhomogeneous
granular structure as suggested in Ref. \cite{german}.
\par
Now consider more closely dependence $L_{\varphi}(T)$ obtained
from analysis of MR in 2D case using Eqs. (\ref{four}) and
(\ref{five}). In the range 3--15 K, dependence $L_{\varphi}\propto
T^{-0.35}$ has been found (Fig.~5). At first glance this seems
unexpected since at low temperatures in 2D systems the phase
relaxation due to EEI should dominates with rate,
$\tau_{ee}^{-1}\propto T$, given by Eq. (\ref{three}), and
$L_{\varphi}\propto T^{-1/2}$ as it is usually observed for
homogeneous enough gold films (see Ref. \cite{belev2,carl} and
references therein). As was indicated above, the diffusion length
of phase relaxation has general temperature dependence
$L_{\varphi}\propto T^{-p/2}$ where index $p$ depends on mechanism
of phase relaxation ($\tau_{\varphi} \propto T^{-p}$). It is known
that in percolating gold films experimental values of the index
$p$ are often found to be smaller than in homogeneous films. For
example, in Refs. \cite{gersh,carl} dependences
$L_{\varphi}\propto T^{-1/3}$ and $L_{\varphi}\propto T^{-0.35}$
were found, respectively, with fitting to Eq. (\ref{four}) for 2D
WL; whereas, $L_{\varphi}\propto T^{-1/2}$ is expected for $p=1$.
In the first case \cite{gersh}, it was attributed to influence of
1D regions in the percolating gold film; whereas, in the second
case \cite{carl}, conception of anomalous electron diffusion for
inhomogeneous films was involved at the explanation. In any case
it is clear that the same phenomenon in the film studied is
determined by inhomogeneity of the film.
\par
The data obtained testify that with increasing temperature the
film becomes more homogeneous in respect to quantum interference
effects: 1D effects in MR disappear above 3~K (Figs. 7 and 8);
$R(T)$ becomes purely logarithmic above 12 K as expected for 2D
systems (Fig. 3). The dependence $L_{\varphi} (T)$ behaves in the
line with this tendency, changing above 15~K to $L_{\varphi}
(T)\propto T^{-1.05}$ (Fig. 5), that corresponds to $p=2.1$. This
agrees with results for disordered but homogeneous enough gold
films \cite{belev2}, where this has been convincingly attributed
to influence of electron-phonon phase relaxation.
\par
In conclusion, we have found that the transport properties of the
inhomogeneous ultra-thin quench-condensed gold film at low
temperature reveal simultaneously indications of the quantum
interference 1D and 2D effects of weak localization and
electron-electron interaction. With increasing temperature the
film behaves as more homogeneous and only 2D effects have been
seen. The observed behavior can be explained by inhomogeneous
electron transport at the threshold of thickness-controlled
metal-insulator transition.
\par
Work was supported in part by program "Structure and Properties of
Nanosystems" (Subsection 1.3.) of National Academy of Sciences of
Ukraine.

\newpage
\centerline{\bf{Figure captions}} \vspace{15pt}

Fig.~1. Temperature behavior of sheet resistance of the gold film
studied measured at different magnitudes of the dc voltage
($U_{appl}$). The $R_{\Box}(T)$ curves are shown for zero magnetic
field (filled symbols) and the field $H=4.2$~T (empty symbols).
\vspace{15pt}

Fig.~2. (Color online) Dependences of resistance on applied
voltage, $U_{appl}$, in low temperature range. The inset shows
$R_{\Box}$ {\it vs} $U_{appl}$ at $T=1,36$~K in the region of low
voltages. \vspace{2pt}

Fig.~3. Temperature dependence of sheet resistance of the gold
film measured at $U_{appl}=0.2$~V in zero magnetic field and field
$H=4.2$~T. The $R(T)$ curves follows the logarithmic law above
$T\approx 8$~K. Below this temperature, deviations, $\Delta R$
from logarithmic growth of resistance with decreasing temperature
are developed. Temperature behavior of these low-temperature
deviations is presented in the inset as $\Delta R$ {\it vs}
$T^{-1/2}$. \vspace{15pt}

Fig.~4. Magnetoresistance curves of the film studied measured at
applied voltage $U_{appl}=0.2$~V for magnetic field normal to the
film. \vspace{15pt}

Fig.~5. (Color online) Calculated temperature dependence of the
phase relaxation length $L_{\varphi} (T)$ in the film obtained
using Eq. {\ref{four} for different magnitudes of $U_{appl}$.
\vspace{15pt}

Fig.~6. Magnetoresistance curve at $T=1.38$~K and
$U_{appl}=0.2$~V. The inset shows enlarged view of low-field part
of the curve.\vspace{15pt}

Fig.~7. Temperature evolution of the low-field anomaly of MR
attributed to 1D effects in transport properties. \vspace{15pt}

Fig.~8. MR vs. square of magnetic field (at $T=1.38$~K,
$U_{appl}=0.2$~V). Inset shows temperature dependence of
saturation MR in the low-field MR anomaly. \vspace{15pt}

Fig.~9. Temperature dependence of Nyquist length $L_N$ calculated
using Eq. (\ref{ten}) at $U_{appl}=3$~V. \vspace{15pt}

\newpage
\begin{figure}[htb]
\centering\includegraphics[width=0.75\linewidth]{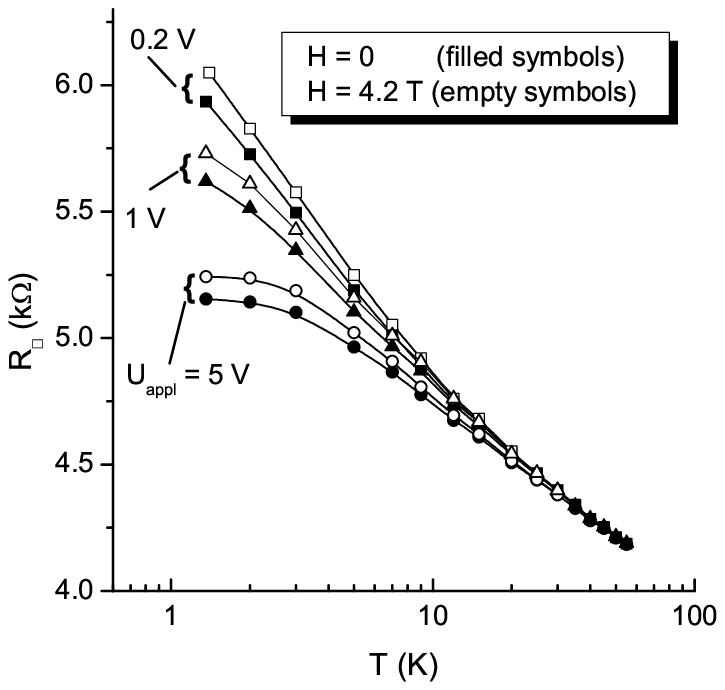}
\centerline{Fig.~1 to paper Beliayev et al.}
\end{figure}

\begin{figure}[tb]
\centering\includegraphics[width=0.8\linewidth]{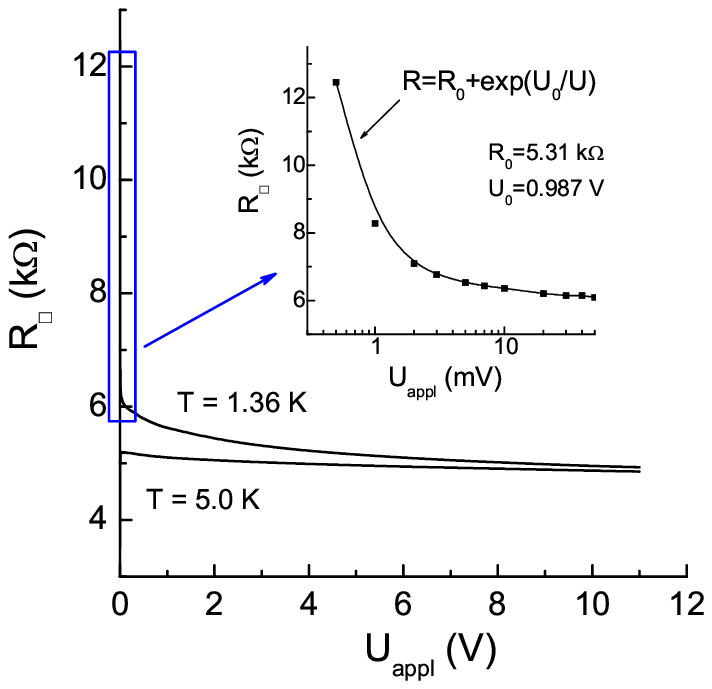}
\centerline{Fig.~2 to paper Beliayev et al. }
\end{figure}

\begin{figure}[htb]
\centering\includegraphics[width=0.9\linewidth]{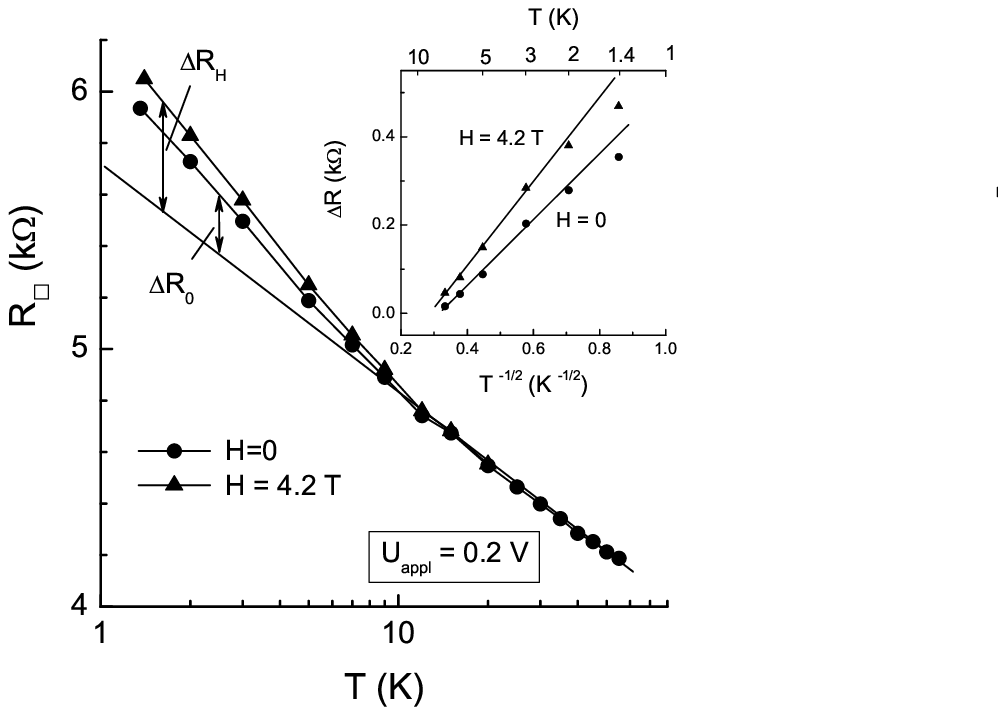}
\centerline{Fig.~3 to paper Beliayev et al.}
\end{figure}

\begin{figure}[htb]
\centering\includegraphics[width=0.75\linewidth]{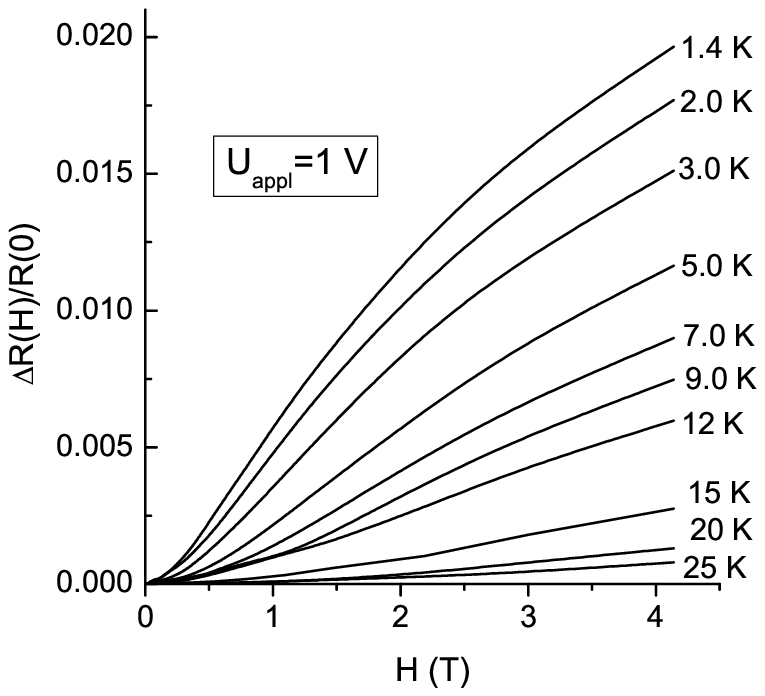}
\centerline{Fig.~4 to paper Beliayev et al.}
\end{figure}

\begin{figure}[tb]
\centering\includegraphics[width=0.75\linewidth]{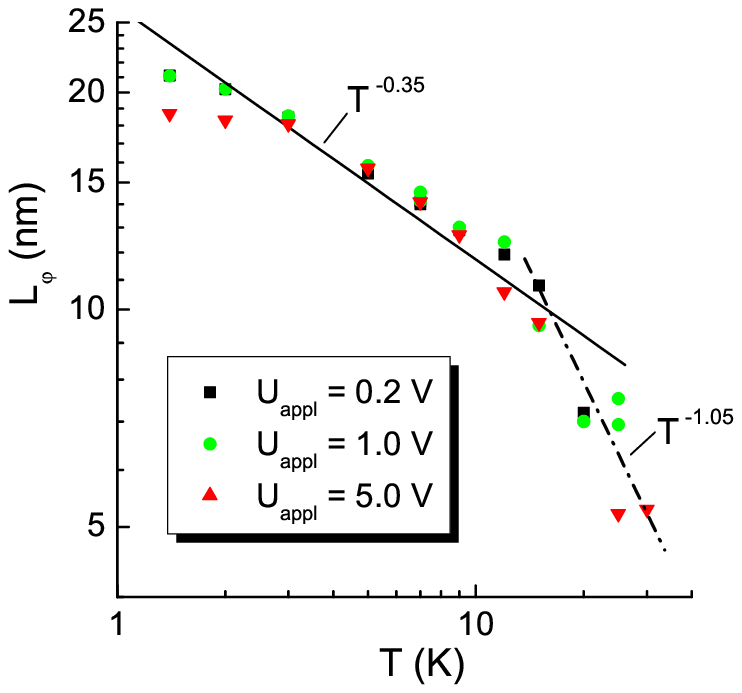}
\centerline{Fig. 5 to paper Beliayev et al. }
\end{figure}

\begin{figure}[htb]
\centering\includegraphics[width=0.75\linewidth]{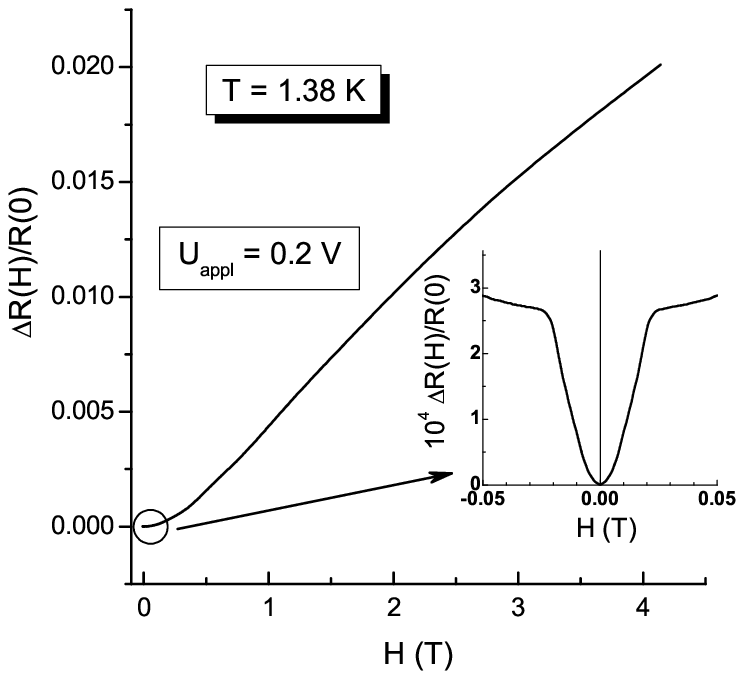}
\centerline{Fig.~6 to paper Beliayev et al.}
\end{figure}

\begin{figure}[tb]
\centering\includegraphics[width=0.75\linewidth]{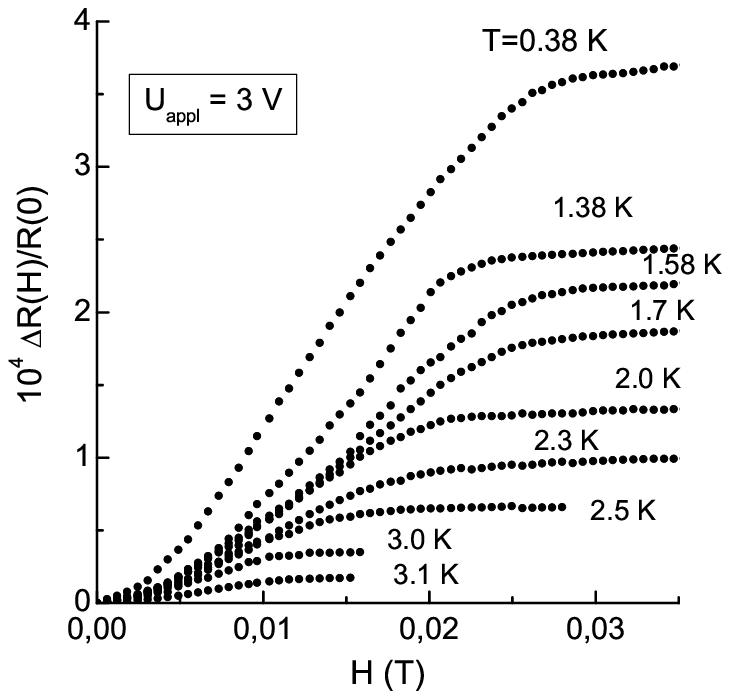}
\centerline{Fig.~7 to paper Beliayev et al.}
\end{figure}

\begin{figure}[tb]
\centering\includegraphics[width=0.75\linewidth]{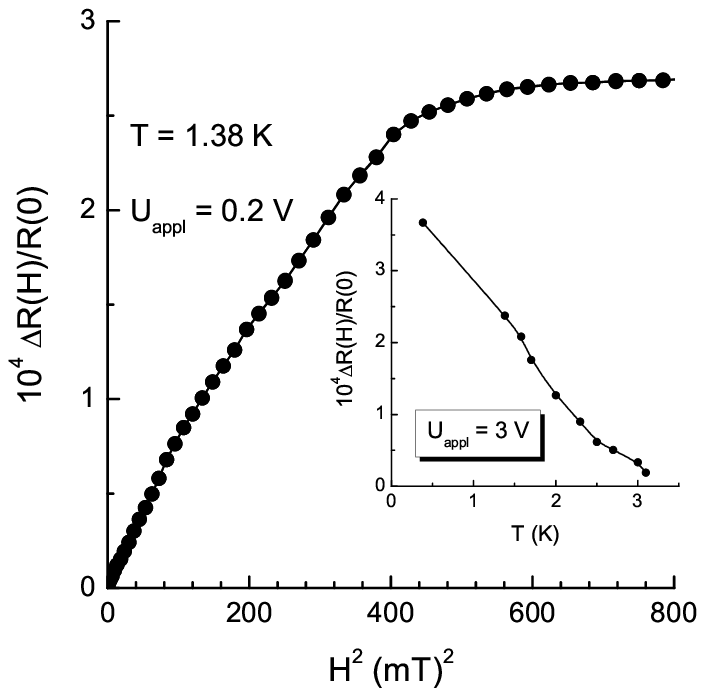}
\centerline{Fig.~8 to paper Beliayev et al.}
\end{figure}

\begin{figure}[tb]
\centering\includegraphics[width=0.75\linewidth]{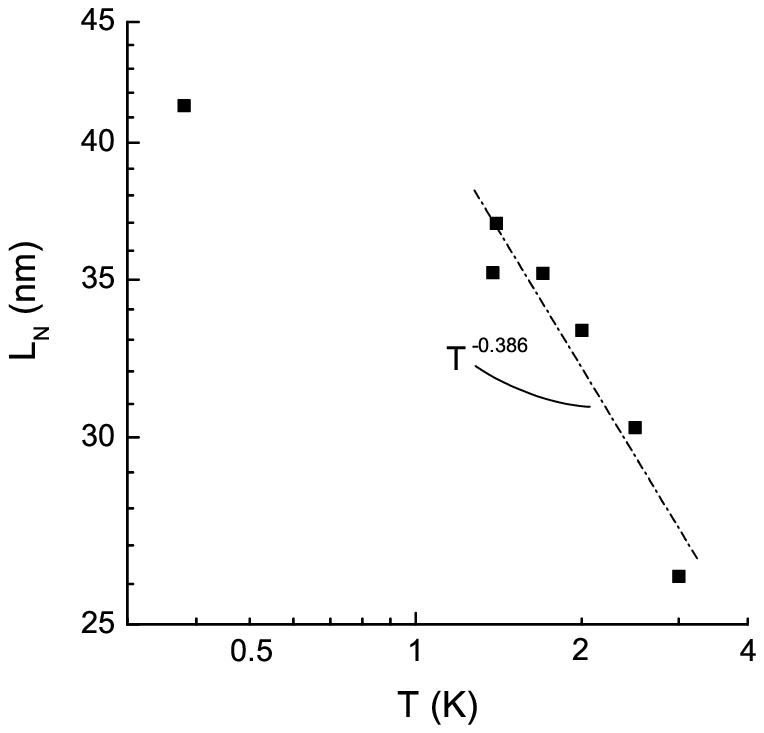}
\centerline{Fig.~9 to paper Beliayev et al.}
\end{figure}

\end{document}